\newcommand{\beq}{\begin{equation}}
\newcommand{\eeq}{\end{equation}}

\newcommand{\id}
 {i\kern.06em\hbox{\raise.25ex\hbox{$/$}\kern-.60em$\partial$}}

\newcommand{\bs}{/\kern-.52em b}
\newcommand{\ds}{/\kern-.52em d}
\newcommand{\qs}{/\kern-.52em s}

\newcommand{\p}{\partial}
\newcommand{\yp}{^{\prime}}

\newcommand{\dd}
{\kern.06em\hbox{\raise.25ex\hbox{$/$}\kern-.60em$\partial$}}

\documentstyle[12pt]{article}
\date{}
\begin{document}
\title{Local observed time and redshift in curved spacetime
 \footnotetext{\# e-mail: sshfeng@yahoo.com }}
\author{{Sze-Shiang Feng $^{1,2,\#}$, }\\
1.{\small {\it CCAST(World Lab.), P.O. Box 8730, Beijing 100080}}\\
2. {\small {\it Department of Astronomy and Applied Physics}}\\
{\small {\it University of Science and Technology of China, 230026, Hefei, China}}}
\maketitle
\newfont{\Bbb}{msbm10 scaled\magstephalf}
\newfont{\frak}{eufm10 scaled\magstephalf}
\newfont{\sfr}{eufm7 scaled\magstephalf}
\baselineskip 0.2in
\begin{center}
\begin{minipage}{135mm}
\vskip 0.3in
\baselineskip 0.2in
\begin{center}{\bf Abstract}\end{center}
  {Using the observed time and spatial intervals defined originally by
 Einstein and the observation frame in the vierbein formalism,
 we propose that in curved spacetime, for a wave received in laboratories, 
the observed frequency is the changing rate of the phase of the 
wave relative to the local observable time scale and the 
momentum the changing rate of the phase relative to the local
 observable spatial length scale. The case of Robertson-Walker 
universe is especially considered.  
\\PACS number(s): 98.60.Eg,98.80.Es,04.20-q   }
\end{minipage}
\end{center}
\vskip 1in
The problem of redshift is very important in cosmology. It is directly related to the estimation of the age of the universe and other observables such as the recession velocities. The standard analysis of redshift of electromagnetic waves, or photons, in static curved spacetime is based on the null geodesics of light and the observable time intervals of the source and the observation point\cite{s1}-\cite{s2}. The standard redshift formula\cite{s2} in the Robertson-Walker universe relates wavelength and the cosmic scale factor. These two cases can be systematically studied by introducing the so-called local oberservation frame\cite{s3}.\\ 
\indent It is well-known that in the vierbein formalism of general relativity there are two kinds of quantities: Riemannian quantities such as the curvature $R_{\mu\nu\alpha\beta}$ and Lorentz quantities such as $R_{abcd}$ which is related to $R_{\mu\nu\alpha\beta}$ in the way
\beq
R_{abcd}=e^{\mu}_ae^{\nu}_be^{\alpha}_ce^{\beta}_dR_{\mu\nu\alpha\beta}
\eeq
Our conventions are as follows
\beq
g_{\mu\nu}=\eta_{ab}e^a_{\mu}e^b_{\nu},\,\,\,\,\, \eta_{ab}={\rm diag}(1,-1,-1,-1)
\eeq
We have to specify which is observable. Since the measurement of space and time intervals is fundamental to those of other quantities, we should study them first. Einstein\cite{s4} pointed out that in a local system, for two neighboring events, the spacetime interval is
\beq
ds^2=g_{\mu\nu}dx^{\mu}dx^{\nu}=\Delta T^2-\Delta L^2
\eeq
where $\Delta L$ is measured directly by a measuring rod and $\Delta T$ by a clock at rest relative to the system :these are the naturally measured lengths and times. \\
\indent An observer is characterized by his position and his velocity in curved spacetime. For a static observer ${\cal O}$ whose velocity is $u^{\mu}=(g_{00}^{-1/2},0,0,0)$. The measured time and space interval is
\cite{s1}
\beq
\Delta T=\frac{g_{\mu 0}}{\sqrt{g_{00}}}dx^{\mu}
\eeq
\beq
\Delta L^2=(\frac{g_{oi}g_{0j}}{g_{00}}-g_{ij})dx^idx^j
\eeq
($i,j=1,2,3$)
This is equivalent to saying that the time rod of ${\cal O}$ is fixed by
\beq
e^0_{\mu}=u_{\mu}
\eeq
This is because the $u_{\mu}=g_{0\mu}/\sqrt{g_{00}}$, so
\beq
e^0_{\mu}dx^{\mu}=\frac{g_{\mu 0}}{\sqrt{g_{00}}}dx^{\mu}
\eeq
So the space interval must be (5). Using ($a^{\yp}=1,2,3$)
\beq
g_{00}=(e^0_0)^2-\eta_{a^{\yp}b^{\yp}}e^{a^{\yp}}_0e^{b^{\yp}}_0
\eeq
we have
\beq
e^{a^{\yp}}_0=0
\eeq
Eqs(6) and (9) determine the time rod of ${\cal O}$; The directions of the spatial rods remain arbitary, i.e., the observer may rotate arbitrarily his apparatus. For a moving observer ${\cal O}^{\yp}(x, u^{\yp})$, the time rod is still fixed by the relation (6)\cite{s3}. Denote the corresponding observation frame as $\bar{e}^a_{\mu}$, then according to special relativity, $\bar{e}^a_{\mu}$ should be related to $e^a_{\mu}$ by a local Lorentz transformation which is determined by the relative Lorentz velocity $u^{\yp a}=u^{\yp\mu}e^a_{\mu}$
\beq
\bar{e}^a_{\mu}(x)=\Lambda^a\,_b(u^{\yp a})e^b_{\mu}(x)
\eeq
where $\Lambda^a\,_c\Lambda^b\,_d\eta_{ab}=\eta_{cd}$. Thus $\Lambda$ is determined up a spatial rotation
\beq
\Lambda^0\,_b=u_b^{\yp}
\eeq
Now the redshift of two light waves in the geometric-optics approximation can be discussed as follows. Consider two static observers $A$ and $B$; two light waves start at $x^{\mu}_A$ and $x^{\mu}_A+dx^{\mu}_A$ to travel from $A$ to $B$. The two worldlines
are null. 
\beq
dt=\frac{-g_{0i}dx^I+\sqrt{(g_{0i}g_{0j}-g_{00}g_{ij})dx^idx^j}}{g_{00}}
\eeq
In general this is a differential equation which can be integrated in some special cases. Once (12) is solved, we can obtain the world points $x^{\mu}_B$ and $x^{\mu}_B+dx^{\mu}_B$ at which the two waves arrive at $B$. Then the redshift is
\beq
\frac{\nu_B}{\nu_A}=\sqrt{\frac{g_{00}(x_B)}{g_{00}(x_A)}}\frac{g_{0\mu}(x_A)dx^{\mu}_A}{g_{0\nu}(x_B)dx^{\nu}_B}=\sqrt{\frac{g_{00}(x_A)}{g_{00}(x_B)}}\frac{dx^0_A}{dx^0_B}
\eeq
For a static metric such as the Schwarzschild metric, it can be easily shown that
$dx^0_A=dx^0_B$, so 
\beq
\frac{\nu_B}{\nu_A}=\sqrt{\frac{g_{00}(x_A)}{g_{00}(x_B)}}
\eeq
For the time-dependent R-W metric of the universe, we have
\beq
\frac{dx^0_A}{dx^0_B}=\frac{R(x^0_A)}{R(x^0_B)}
\eeq
So the redshift given by
\beq
\frac{\nu_B}{\nu_A}=\frac{R(x^0_A)}{R(x^0_B)}
\eeq
So in the geometric-optics limit, the two cases can be systematically discussed.   \\
\indent In this letter we take into account the diffraction effect referring to the observable time interval (4). We consider the redshift of Klein-Gordon field $\phi$ in curved spacetime. Note that in flat sapcetime, the free particle is represented by a plane wave $e^{-ik_{\mu}x^{\mu}}$. The frequency of the wave observed at ${\bf x}$ is the time-changing rate of the phase and the momentum is the spatial-changing
rate, i.e., the gradient of the phase.  In curved spacetime, suppose that the solution of the covariant Klein-Gordon equation can be expressed as $\Phi=e^{-iS(x)}$ and we can search for the function $S(x)$ satisfying the two equations
\beq
\nabla_{\mu}S(x)\nabla^{\mu}S(x)=m^2,\,\,\,\,\, \nabla_{\mu}\nabla^{\mu}S(x)=0
\eeq
It can be checked that the Klein-Gordon equation 
\beq
(\nabla^{\mu}\nabla_{\mu}+m^2)\phi=0
\eeq
is satisfied. Then we can identify the corresponding four-momentum as
\beq
p_a(x)=e^{\mu}_a(x)\nabla_{\mu}S(x)=mu_a
\eeq
because we have identically
\beq
u^au_a=1,\,\,\,\,\, \nabla_{\mu}u^{\mu}=0
\eeq
From $u_{\mu}=\nabla_{\mu}S(x)$ we have 
\beq
\nabla_{\lambda}(\nabla^{\mu}S\nabla_{\mu}S)=2\nabla^{\mu}S\nabla_{\lambda}\nabla_{\mu}S=2\nabla^{\mu}S\nabla_{\mu}\nabla_{\lambda}S=0
\eeq
which is just the geodesic equation. The first one in (17) is just the Jaccobi equation of a classical particle in a gravitational field.  As argued in \cite{s3} it is $p_a$ rather than $p_{\mu}:=\nabla_{\mu}S$ which is observable. But in general, the solution is of the form $\Phi=|\Phi| e^{-iS(x)}$, and even if $|\Phi|=1$, we are unlikely to obtain $S$ satisfying both equations of (17). Yet, we are still able to recognize the momentum and frequency of the wave at a point $x^{\mu}$. Our main argument in this letter is that {\it As in the flat spacetime, the frequency is still the changing rate of the phase relative to the local observable time scale and the momentum the changing rate of the phase relative to the local observable spatial length scale.} \\
\indent We now consider a Klein-Gordon field in the R-W spacetime. The equation of motion is 
\beq
\Box\Phi+\xi R\Phi=0
\eeq
where $\xi$ is the coupling of the field $\Phi$ with the scalar curvature $R$. The R-W metric in the conformal coordinates is\cite{s5}
\beq
g_{\mu\nu}=C(\eta)(1,-h_{ij}({\bf x}))
\eeq
where $C(\eta)=a^2(t), a(t)$ is the cosmic scale factor. $\eta$ is the conformal time, $d\eta=\frac{dt}{a(t)}$, and $h_{ij}({\bf x})$ is defined as
\beq
\sum^3_{i,j}h_{ij}dx^idx^j=(1-\kappa r^2)^{-1} dr^2+r^2(d\theta^2+\sin^2\theta d\phi^2)
\eeq
with $\kappa=0,1,-1$ for flat, positive, or negative curved spatial sections, respectively. The homogeneity and isotropy of the metric allows a separation of the time and spatial dependence of (22), so we make the general ansatz
\cite{s5}\cite{s6}
\beq
\Phi_{\underline{k}}={\cal Y}_{\underline{k}}({\bf x})C^{-1/2}(\eta)f_{\underline{k}}(\eta)
\eeq
with ${\bf x}=(r,\theta,\phi)$ and ${\cal Y}_{\underline{k}}({\bf x})$ an eigenfunction of the three spatial Laplacian with eigenvalue $-\lambda^2$\cite{s7}
\beq
h^{-1/2}\p_i[h^{1/2}h^{ij}\p_j{\cal Y}_{\underline{k}}({\bf x})]=-\lambda^2 {\cal Y}_{\underline{k}}({\bf x})
\eeq
For $\kappa=0$,
\beq
{\cal Y}_{\underline{k}}({\bf x})=(2\pi)^{-3/2}e^{-i{\bf k}\cdot{\bf x}},\,\,\,\,\,\,\,
\underline{k}={\bf k}=(k_1,k_2,k_3), \,\,\,\,(-\infty<k_i<\infty)
\eeq
For $\kappa=1$, if the three-metric is written as
\beq
\sum h_{ij}dx^idx^j=d\omega^2+\sin^2\omega d\alpha^2+\cos^2\omega d\beta^2
\eeq
then
\beq
\kappa=1: {\cal Y}_{\underline{k}}({\bf x})=d^{l/2}_{nm}(\omega)e^{in\alpha}e^{im\beta},\,\,\,\,\,\,\underline{k}=(l.m.n),\,\,\,\, (l=0,1,...,; n,m=-\frac{1}{2}l, -\frac{1}{2}l+1,..., \frac{1}{2}l)
\eeq
where $d^{l/2}_{nm}$ is a representation function of the group SU(2). On the other hand, when the isotropy of the space about one point is exhibited by writing
\beq
\sum h_{ij}dx^idx^j=d\chi^2+\sin^2\chi(d\theta^2+\sin^2\theta d\phi^2)
\eeq
, we have
\beq
\kappa=1: {\cal Y}_{\underline{k}}({\bf x})=\Pi^{(+)}_{lJ}(\chi)Y^M_J(\theta,\phi),\,\,\,\underline{k}=(l,J,M),\,\,\,
(l=0,1,...; J=0,1,...,l; M=-J,-J+1,...,J)
\eeq
where the $Y^M_J$ are the usual spherical harmonics with an appropriate phase.
In the case of open three-space, $\kappa=-1$,
\beq
\sum h_{ij}dx^idx^j=d\chi^2+{\rm sinh}^2\chi(d\theta^2+\sin^2\theta d\phi^2)
\eeq
The eigenfunctions are obtained by replacing $\chi$ by $i\chi$ and $l+1$ by $iq$ in the formulae for $\kappa=1$. However, $q$ is now a continuous variable and $J$ can be arbitrarily large:
\beq 
\kappa=-1: {\cal Y}_{\underline{k}}({\bf x})=\Pi^{(-)}_{qJ}(\chi)Y^M_J(\theta,\phi),\,\,\,\,\underline{k}=(q,J,M),\,\,
(0<q<\infty; J=0,1,...; M=-J,-J+1,...,J)
\eeq
The eigenvalue $-\lambda^2$ is determined by
\beq
\lambda^2=\Bigl\{
\matrix{|{\bf k}|^2 & {\rm if}  \kappa=0\cr
l(l+2)    & {\rm if}  \kappa=1\cr
q^2+1     &  {\rm if} \kappa=-1\cr}
\eeq
The functions $\Pi^{(-)}$ are defined by\cite{s8}
\beq
\Pi^{(-)} _{kJ}(x)=\{\frac{\pi}{2}k^2(k^2+1)...[k^2+(2+J)^2]\}^{-1/2}\sinh x(\frac{d}{d\cosh x})^{1+J}\cos kx
\eeq
while $\Pi^{(+)}_{kJ}$ can be obtained from $\Pi^{(-)}_{kJ}$ by replacing $k$ by $-ik$ and $x$ by $-ix$ in the latter\cite{s9}. So the phase of ${\cal Y}_{\underline{k}}({\bf x})$ is $ e^{i{\bf k}\cdot{\bf x}}$ for $\kappa=0$ and $e^{iM\phi}$ for $\kappa=\pm 1$. With the separation (25),$f_{\underline{k}}$ satisfies
\beq
\ddot{f}_{\underline{k}}(\eta)+[\lambda^2+6\xi\kappa]f_{\underline{k}}(\eta)
+(3\xi-\frac{1}{2})[\frac{\ddot{C}}{C}-\frac{1}{2}(\frac{\dot{C}}{C})^2]
f_{\underline{k}}(\eta)=0
\eeq
where the overdot denotes derivative with respect to conformal time. Both the $\eta$-dependence and the ${\bf x}$-dependence are necessary in order to analyze the frequency and the momentum. \\
\indent  For a generic power-law expansion $a(t)\sim t^p$, (36) reduces to the Bessel's equation\cite{s10}
\beq
\ddot{f}_{\underline{k}}(\eta)+[\beta^2-\frac{\nu^2-\frac{1}{4}}{\eta^2}]
f_{\underline{k}}(\eta)=0
\eeq
where
\beq
\beta^2=\lambda^2+6\xi\kappa, \,\,\,\,\,
\nu^2(\xi,p)=\frac{1}{4}-(6\xi-1)\frac{p(2p-1)}{(p-1)^ 2}
\eeq
So the $f_{\underline{k}}(\eta)$ is a combination of Hankel functions
\beq
f_{\underline{k}}(\eta)=\eta^{1/2}[\alpha_1({\underline{k}})H^{(1)}_\nu(\beta\eta)+
\alpha_2({\underline{k}})H^{(2)}_\nu(\beta\eta)]
\eeq
for all $p\not=1$. For both $p=1$ (curvature-dominated expansion) and $1/2$
(radiation-dominated expansion), eq(36) permits ordinary plane-wave solutions.
As argued in \cite{s10},$ H^{(1)}_\nu(\beta\eta)$ corresponds to the positive-frequency modes appropriate for R-W background. The time-dependent part of the phase of $\Phi$ is completely contained in the phase of $ H^{(1)}_\nu(\beta\eta)$
which is easy to identify\cite{s10}
\beq
 H^{(1)}_\nu(\beta\eta)\equiv A(\beta\eta,\nu)e^{-iS(\beta\eta,\nu)}
\eeq
where $A$ and $S$ are the real valued amplitude and phase functions. $S$ is found to be
\beq
S(\beta\eta,\nu)={\rm arctan}\frac{{\rm cot}(\pi\nu)J_{\nu}(\beta\eta)-{\rm csc}
(\pi\nu)J_{-\nu}(\beta\eta)}{J_{\nu}(\beta\eta)}
\eeq
\beq
S(\beta\eta,\nu)={\rm arctan}
\frac{{\rm Im}[e^{-i\pi\nu}J_{\nu}(\beta\eta)-J_{-\nu}(\beta\eta)]}{{\rm Re}[e^{-i\pi\nu}J_{\nu}(\beta\eta)-J_{-\nu}(\beta\eta)]}
\eeq
for real and imaginary $\nu$ respectively. For the R-W metric, the time-rod for a static observer is fixed by $e^{\mu}_0(\eta,{\bf x})=\delta^{\mu}_0\frac{1}{C^{1/2}(\eta)}$ in the conformal coordinates. Therefore
the frequency observed at point $(\eta,{\bf x})$ is
\beq
\omega_{\underline{k}}(\eta,{\bf x})=e^{\mu}_0 \p_{\mu}S=C^{-1/2}(\eta)\frac{\p S}{\p\eta}
\eeq
Hence in the comoving coordinates $(t,{\bf x})$, the wave length observed is given by
\beq
\lambda(t)=\frac{2\pi a(t)}{(\p S/\p\eta)(t)}
\eeq
The redshift is then
\beq
Z=\frac{\lambda_0}{\lambda_1}-1=\frac{a(t_0)}{a(t_1)}\frac{(\p S/\p\eta)(t_1)}
{(\p S/\p\eta)(t_0)}-1
\eeq
This formula was first given in \cite{s10}.  \\
\indent To investigate the momentum of the corresponding quanta observed at
$(\eta,{\bf x})$, we first consider the case $\kappa=0$. From (27) it is seen that for the mode $\Phi_{\underline{k}}$, the momentum projected onto the spatial axis is
\beq
{\bf p}(\eta,{\bf x})= C^{-1/2}(\eta){\bf k}
\eeq
So in general, the observed frequency and the momentum do not satisfy the relativistic relation of a massless particle $\omega^2_{\bf k}={\bf p}^2$. In the short wave limit,
$\beta\rightarrow\infty$, the phase has the asymptotic behavior
\beq
S(\beta\eta,\nu)\sim \beta\eta-\frac{1}{2}\pi\nu(\xi,p)-\frac{1}{4}\pi
\eeq
This limit is the geometric-optics limit
\beq
\omega_{\bf k}(\eta,{\bf x})=C^{-1/2}(\eta)\beta= C^{-1/2}(\eta)|{\bf k}| 
\eeq
Therefore, in this limit the relativistic relation is satisfied. This is reasonable since in the short-wave limit, it is the field quanta rather than  wave is observed and in the long-wave case, the field is more wave-like than particle-like. The deviation of the frequency-momentum from $\omega^2_{\bf k}={\bf p}^2$ shows the effect of diffraction. For $\kappa=\pm 1$ in the coordinates $(r,\theta,\phi)$, we can  simply choose the inverse vierbein as 
\beq
e^{\mu}_a=C^{-1/2}(\eta)(1,-\frac{1}{\sqrt{1-\kappa r^2}}, -\frac{1}{r},-\frac{1}{r\sin\theta})
\eeq
we have the three-momentum for the mode $\Phi_{\underline{k}}$ observed by a static observer
\beq
{\bf p}(\eta,{\bf x})=(0,0,C^{-1/2}(\eta) \frac{1}{r\sin\theta}\p_{\phi}(M\phi))= (0,0,C^{-1/2}(\eta) \frac{1}{r\sin\theta}M)
\eeq
Bearing in mind that $M$ is roughly the $z$-component of the angular-momentum, i.e., $M\sim l_z\sim p_{\phi}r\sin\theta$ ($p_{\phi}$ is projection of the momentum along the $\phi$-direction) the  and the $e^{\phi}_{a^{\yp}}$ denotes the $\phi$-rod of the observer, hence our conclusion (50) is reasonable. As in the case for $\kappa=0$, in general the relation $\omega^2={\bf p}^2$ is not satisfied either. The deviation exhibits the affect of diffraction. \\
\indent In this letter we discussed the observed frequency and momentum of a field in curved spacetime with Einstein's local observable spacetime intervals. In particular, we discussed the $\Phi$-field in the R-W spacetime. The discussion obviously applies to a vector field. Our conclusions for the $\Phi$-field are physically reasonable. Since the R-W spacetime covers a family of spacetimes such as de Sitter spacetime, our discussion applies also to those spacetimes. It is expected that our definition of observed frequency and momentum can find some application in both cosmology and particle physics.
\vskip 0.3in
\underline{Acknowledgement}Work supported by the NSF of China under Grant No. 19805004.

\end{document}